\documentclass{Interspeech}

\interspeechcameraready

\title{TF-Mamba: A Time-Frequency Network for Sound Source Localization}

\author[affiliation={1,2}]{Yang}{Xiao}
\author[affiliation={2}]{Rohan Kumar}{Das}
\affiliation{}{The University of Melbourne}{Australia}
\affiliation{}{Fortemedia Singapore}{Singapore}

\email{yxiao9550@student.unimelb.edu.au, rohankd@fortemedia.com}

\usepackage{comment}
\usepackage{multirow} 
\usepackage{cite}
\usepackage{amsmath,amssymb,amsfonts}
\usepackage{algorithmic}
\usepackage{graphicx}
\usepackage{textcomp}
\usepackage[table,xcdraw]{xcolor}
\usepackage{balance}
\usepackage{pifont}% http://ctan.org/pkg/pifont
\newcommand{\cmark}{\ding{51}}%
\newcommand{\xmark}{\ding{55}}%
\begin{document}

\maketitle

% the abstract here must exactly match the abstract entered into the paper submission system
\begin{abstract}
    Sound source localization (SSL) determines the position of sound sources using multi-channel audio data. It is commonly used to improve speech enhancement and separation. Extracting spatial features is crucial for SSL, especially in challenging acoustic environments. Recently, a novel structure referred to as Mamba demonstrated notable performance across various sequence-based modalities. This study introduces the Mamba for SSL tasks. We consider the Mamba-based model to analyze spatial features from speech signals by fusing both time and frequency features, and we develop an SSL system called TF-Mamba. This system integrates time and frequency fusion, with Bidirectional Mamba managing both time-wise and frequency-wise processing. We conduct the experiments on the simulated and real datasets. Experiments show that TF-Mamba significantly outperforms other advanced methods. The code will be publicly released in due course.
\end{abstract}

\section{Introduction}
\label{sec:intro}
Sound source localization (SSL) \cite{ssl1} aims to automatically identify the origins of sound by analyzing signals captured through a microphone array. The core functionality of SSL systems is to calculate the angles at which sounds reach the microphones. This spatial information is essential for many speech-based applications, including speech separation \cite{ss,tse,xiaoicassp}, speech recognition \cite{asr}, and speech enhancement \cite{se}. Accurate sound localization can assist these applications to improve their performance. As a result, developing more advanced SSL techniques has become a key focus in research, aiming for greater robustness and adaptability in the real world.

Traditional SSL methods estimate spatial features linked to direct-path signal propagation to map features to source locations. Common spatial features include time delay, inter-channel phase/level difference (IPD/ILD)~\cite{itd}, and relative transfer function (RTF)~\cite{rtf}. For example, the steered response power (SRP) based methods, particularly SRP-PHAT~\cite{srp}, which is obtained from the SRP by applying phase transform whitening, have been foundational due to their simplicity and versatility. These spatial features are easy to estimate in ideal conditions but become challenging in real-world scenarios with noise, reverberation, and multiple moving sources. Because the noise and overlapping speech cause distortions, while reverberation masks or colors the signal. Additionally, moving sources create time-varying spatial cues, also leading to a significant decline in localization accuracy.

In recent years, deep learning methods for localization have gained more attention~\cite{fu2,fu3,cdoa} than traditional methods. These methods approach the task as either feature/location regression or location classification. Common architectures for moving SSL include convolutional neural networks (CNNs)~\cite{cnn1,cnn2} and recurrent neural networks (RNNs)~\cite{cnn3,cnn4}. CNNs extract local spatial features, while RNNs capture long-term temporal context. Deep learning models can take input at the signal level (e.g., time-domain signals, spectrograms) or feature level (e.g., inter-channel phase difference, spatial spectrum). Inspired by speech enhancement, the recent FN-SSL~\cite{fu1} framework builds on previous research by effectively estimating the localization of the sound source in both the narrow-band (frequency) and full-band (timeframe) domains. Long short-term memory (LSTM) layers are used to process each band, demonstrating the state-of-the-art performance.

Although RNN models like FN-SSL have shown promising results, a new architecture based on the neural state space model (SSM) called Mamba \cite{ssm1, ssm2} has emerged and achieves superior performance to state-of-the-art models in various tasks \cite{visionmamba, jamba, mambainspeech}. In speech processing, there have been attempts to replace transformers and RNNs with Mamba for tasks such as speech separation and speech enhancement \cite{spmamba, xlsrmamba, semamba}. As~\cite{mamba} state, Mamba outperforms RNNs in modeling very long-range dependencies and is recognized for its efficient use of computational resources which is suitable for SSL applications.

% In this work, we {\it introduce Mamba to SSL} by proposing a novel architecture referred to as TF-Mamba. It is built on the robust FN-SSL framework, which uses LSTM for time and frequency feature fusion. By replacing the LSTM blocks in FN-SSL with bidirectional Mamba (BiMamba) blocks, TF-Mamba aims to improve audio sequence context modeling while maintaining linear complexity with sequence length. We consider both simulated and LOCATA datasets for our studies to evaluate the effectiveness of proposed TF-Mamba for real-world scenarios. To the best of our knowledge, this is the first attempt to use a state-space-based model for SSL. 
In this work, we {\it introduce Mamba to SSL} by developing a novel architecture called TF-Mamba. Our goal is to enhance SSL by utilizing Mamba's capability to efficiently model long-range dependencies. Additionally, we apply Mamba to both temporal and frequency dimensions, creating a dual-dimension approach that improves feature extraction and better captures spatial and localization cues compared to baseline models. By utilizing bidirectional Mamba (BiMamba) blocks, TF-Mamba enhances audio sequence context modeling while maintaining computational complexity. We evaluate TF-Mamba using both simulated data and the LOCATA dataset, demonstrating its effectiveness in real-world scenarios. To our knowledge, this is the first study to apply a state-space-based model to SSL, showcasing the novelty and significance of our approach.
\begin{figure*}[t]
\centering  
\includegraphics[width=\linewidth]{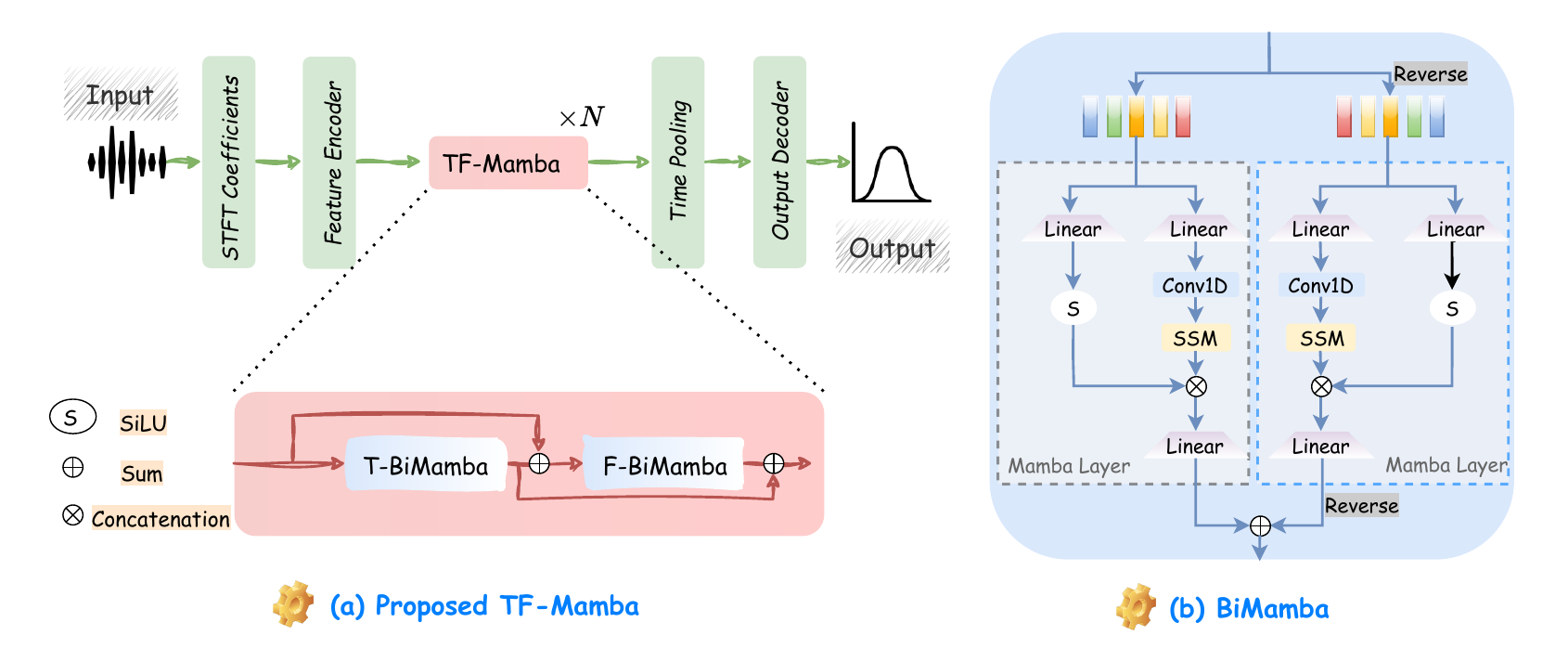}
\vspace{-4mm}
\caption{Architecture of (a) the proposed TF-Mamba network and (b) Bidirectional Mamba (BiMamba) layer. Each TF-Mamba block includes a temporal Mamba (T-BiMamba) and frequency Mamba (F-BiMamba) layer, with skip connections to prevent information loss. ``S" denotes the SiLU activation and ``Linear" indicates linear projection.}
\label{fg1}
\vspace{-4mm}
\end{figure*}

\section{Related Works}
\label{sec:related}
\subsection{Mamba for Speech}
Mamba~\cite{mamba} has demonstrated transformer-level performance across various sequence-based modalities, including audio and speech, which are naturally sequential in waveform or spectrogram forms. Early works applied Mamba to single tasks such as speech enhancement~\cite{semamba, se1}, speech separation~\cite{spmamba}, and audio detection and classification~\cite{audiomamba,rawbmamba}. Additionally, some studies explored self-supervised audio transformers trained with masked spectrogram modeling. The most comprehensive studies to date examine Mamba's applications in speech enhancement, recognition, synthesis, understanding, and summarization. However, effective model design using SSMs for deep SSL remains unexplored.
% https://arxiv.org/pdf/2407.09732, https://arxiv.org/pdf/2405.12609
\subsection{Deep Sound Source Localization}
In recent years, significant progress has been made in SSL using neural networks. Cross3D~\cite{mv} uses the SRP-PHAT spatial spectrum as input and employs a causal 3D CNN network for tracking moving sound sources. Similarly, SE-ResNet~\cite{seresnet} utilizes a squeeze-and-excitation residual network as an encoder and a gated recurrent unit (GRU) network as a decoder. While SELDnet~\cite{seldnet} takes frequency-normalized inter-channel phase difference (IPD) concatenated with the magnitude spectrum as input, SALSA-Lite~\cite{salsa}, similar to SELDnet, uses frequency-normalized IPD with the magnitude spectrum as input and a ResNet-GRU network. With advanced performance, the FN-SSL~\cite{fu1} also uses direct path IPD as input and leverages dedicated full-band and narrow-band LSTM layers to exploit inter-band dependencies and inter-channel information. 
% https://arxiv.org/pdf/2405.07021, https://arxiv.org/pdf/2305.19610

\section{Proposed TF-Mamba}
\label{sec:method}
\subsection{BiMamba}

The structured SSMs~\cite{ssm1,ssm2} efficiently handle long-dependent sequences with low computation and memory needs, serving as a substitute for transformers or RNNs. Mamba \cite{mamba} enhances SSM by introducing an input-dependent selection mechanism for efficient information filtering and a hardware-aware algorithm that scales linearly with sequence length, enabling faster computation. Mamba’s architecture, combining SSM blocks with linear layers, is simpler but achieves state-of-the-art performance across various long-sequence tasks, offering significant computational efficiency during both training and inference.

The core of Mamba is a linear selective SSM. The SSM transforms an input \(x_t\) into an output \(y_t\) using a hidden state \(h_t\) at each step \(t\) in the sequence. It includes three learnable matrices: \(A\) for state transitions, \(B\) for inputs, and \(C\) for outputs. A discretization process is applied to continuous-time SSMs for the integration into practical deep neural architectures. We convert the continuous matrices \(A\) and \(B\) into discrete forms, \(\tilde{A}\) and \(\tilde{B}\). This step allows the model to work with discrete-time signals. After this, the SSM follows the equation shown below:
 
\begingroup
\vspace{-2mm}
\begin{equation}
h_t = \tilde{A} h_{t-1} + \tilde{B} x_t, \quad y_t = C h_t
\end{equation}
\vspace{-4mm}
\endgroup

Because of its linear structure, the full output sequence \(y\) of length \(\mathcal{L}\) can be written as a convolution between the input sequence \(x\) and a kernel \(\mathcal{K}\). Each element of \(\mathcal{K}\) shows how an input at one time step affects future outputs through the state transitions. This kernel summarizes the model’s temporal behavior and allows efficient processing of long sequences. This makes structured state-space models useful for SSL.

\begingroup
\vspace{-4mm}
\begin{equation}
    \quad y = x \ast \mathcal{K}, {\text {~where~}} \mathcal{K} = (C\tilde{B}, C\tilde{A}\tilde{B}, \ldots, C \tilde{A}^{\mathcal{L}-1}\tilde{B}) 
\end{equation}
\vspace{-4mm}
\endgroup

However, the matrices in the model depend on the input \(x_t\), making them input-selective. Because of this, the convolution cannot be computed directly. Instead, it is solved using a parallel scan algorithm. In the Mamba model, the SSM is placed between two gated linear layers. These added components improve performance by introducing selective features, but the model remains unidirectional, like a traditional RNN.
While unidirectionality is acceptable in large language models that use autoregressive training, it becomes a problem in non-autoregressive speech models. For the SSL task, models must capture both past and future context. Therefore, a non-causal module, such as attention, is needed. Finding an effective way to add this capability is an important challenge.

We use the bidirectional Mamba (BiMamba) proposed in~\cite{semamba,mambainspeech}, where two SSMs and causal convolutions run in parallel: one processes the original sequence, and the other processes the reversed sequence. The outputs from both SSMs are averaged to incorporate information from both directions.

\subsection{Network Architecture}
The structure of the proposed TF-Mamba network is illustrated in Figure~\ref{fg1} (a). The input to the network consists of the real and imaginary parts of the STFT coefficients, resulting in an input channel count that is twice the number of microphones. The model has three modules as follows: \subsubsection{Feature Encoder} The inputs are first processed by a feature encoder, which consists of a dilated DenseNet~\cite{densenet} core flanked by two convolutional layers to enhance its spectral properties. The output from the feature encoder is then transformed through the Time-Frequency Mamba block. 
% Through alternating temporal and frequency Mamba layers, one temporal Mamba layer and one frequency Mamba layer form a TF-Mamba block.  

\subsubsection{Time-Frequency Mamba}
The Time-Frequency Mamba  (TF-Mamba) block consists of one temporal Mamba layer and one frequency Mamba layer. Figure~\ref{fg1} (a) illustrates a network with a TF-Mamba block, with additional blocks easily added by repeating the block multiple ($N$) times. Figure~\ref{fg1} (b) illustrates the BiMamba layer. 

{\bf The temporal Mamba layers} process time frames independently, with all frames sharing the same network parameters. The input is a sequence along the frequency axis of a single time frame, allowing the temporal Mamba layers to learn inter-frequency dependencies related to spatial and localization cues. These layers do not capture temporal information, which is instead handled by the subsequent frequency Mamba layers.

{\bf The frequency Mamba layers} process frequencies independently, with all frequencies also sharing parameters. Differing from the temporal Mamba, the input is a sequence along the time axis of a single frequency, where the frequency Mamba layers estimate direct-path localization features. Estimating these features frequency-wise has been extensively studied in conventional methods such as in~\cite{fu1}. The proposed frequency Mamba layers focus on exploiting this inter-channel information. Additionally, since it is time-varying for moving sound sources, the frequency Mamba layers also learn the temporal evolution of the direct-path localization features.

Mamba is useful because it efficiently models long-range dependencies while maintaining computational efficiency. The ability of Mamba to handle both temporal and frequency-specific features in a structured manner makes it ideal for capturing complex relationships in audio data, such as the time-varying nature of localization cues, which are crucial for accurate SSL. However, the temporal Mamba and frequency Mamba layers are designed to focus on their respective types of information. Therefore the pieces of information from the frequency Mamba layer can be lost after passing through a temporal Mamba layer, and vice versa. To prevent this information loss, we add skip connections. As shown in Figure~\ref{fg1}, the output of each temporal Mamba layer is added to the input of the following temporal Mamba layer. Similarly, the output of each frequency Mamba layer is added to the input of the next frequency Mamba layer. Further, each Mamba layer inside BiMamba also has the skip connection.

\subsubsection{Output Decoder} The output from the TF-Mamba blocks is then passed through an average pooling module to reduce the frame rate and is subsequently fed into a fully connected layer with $\tanh$ activation, which converts the output to the desired dimension. We employ a spatial spectrum prediction module for DOA estimation as in~\cite{ssnet1}. This system learns complex relationships from the audio data to predict the potential origin of sound at every degree within a half-circle of the linear microphone array. The resulting spatial spectrum is an 181-point map. We train our model by minimizing the mean squared error (MSE) loss between the model's output spatial spectrum and the ground truth.
\section{Experiment Setting}
\label{sec:exp}
\subsection{Datasets}
We evaluated the proposed method on both simulated and real-world datasets, using two microphones to localize the direction of arrival (DOA) within a 180-degree azimuth range.

\textbf{Simulated dataset:} Microphone signals are obtained by convolving room impulse responses (RIRs) with speech source signals~\cite{cdoa}. The speech signals are randomly selected from the training, development, and test sets of the LibriSpeech corpus~\cite{librispeech}, and then used for model training, validation, and testing, respectively. RIRs are generated using the gpuRIR~\cite{gpurir}. We followed the settings from~\cite{fu1} to obtain the simulated dataset with moving sources. The room reverberation time (RT60) is randomly set between 0.2 and 0.6 seconds, and the room size ranges from \(4\times2\times2\) m to \(10\times8\times5\) m. The moving trajectories of speech sources are generated according to ~\cite{mv}, with each source maintaining a fixed height. Two microphones, placed 8 cm apart, are randomly positioned in the room on the same horizontal plane as the sound source. This dataset includes 1,511 audio files, which amount to approximately 4.96 hours of speech data. Based on the RIRs, we generate the mixture data for each original single-channel speech data for every 5 degrees from 0 to 180 degrees, which expands the audio files by 37 times. We then use white, babble, and factory noises from the NOISEX-92~\cite{NOISEX92} dataset as noise sources and create a diffuse sound field described in~\cite{generating}. The generated diffuse noise signals are added to the clean signals with a signal-to-noise ratio (SNR) between -10 and 10 dB for noise robustness.

\textbf{Real-world dataset:} We also tested our proposed method on tasks 3 and 5 of the LOCATA dataset~\cite{locata} following the setting in~\cite{fu1}. The room size was \(7.1\times9.8\times3\) m, with a reverberation time of 0.55 s. We used microphones 6 and 9 from the DICIT array~\cite{woz}, which have the same configuration as the simulated microphone array. The models trained on the simulated dataset were directly tested on the LOCATA dataset.

\subsection{Implementation details and metrics}

For training each task with our proposed TF-Mamba model, we use 100 epochs with a 20-epoch early stopping criterion. We utilize the AdamW optimizer with a learning rate of 0.001. Additionally, we employ StepLR, to adjust the learning rate during training. For STFT computations, the frame size is set to 32 milliseconds, and the hop size to 10 milliseconds. The frequency range for the STFT is between 100 Hz and 8000 Hz, with an n\_fft value of 512, which determines the length of the windowed signal after zero-padding. 

We use two primary metrics: mean absolute error (MAE) and accuracy (ACC). MAE quantifies the average magnitude of prediction errors, disregarding their direction. ACC, on the other hand, measures the percentage of predictions that are exactly correct or within a specified tolerance. We assess accuracy at two tolerance levels: within a \(10^\circ\) and \(15^\circ\) error margin. These tolerance levels provide insight into the model's precision as well as its practical effectiveness.
% Please add the following required packages to your document preamble:
% \usepackage{graphicx}

\subsection{Reference baselines}
We compared the following five moving SSL methods: Cross3D~\cite{mv}, SELDnet~\cite{seldnet}, SE-Resnet~\cite{seresnet}, SALSA-Lite~\cite{salsa}, and FN-SSL~\cite{fu1}. It is noted that the SELDnet, SALSA-Lite, and SE-ResNet are models designed for joint sound event detection and localization; hence, we only utilize their localization branches for comparison. All these methods are reproduced and trained on the same dataset as our proposed method. We adjust the time pooling kernel size in each method to achieve the same output frame rate across all models.

\begin{table}[t]
\centering
\caption{Ablation studies of TF-Mamba on simulated data. ``Expand" denotes the expand factor for the Mamba layer.}
\vspace{-2mm}
\label{tab:abl-table}
\resizebox{\linewidth}{!}{%
\begin{tabular}{l|c|c|c|c}
\toprule
\textbf{Components}            & \textbf{Params{[}M{]}} $\downarrow$ & \textbf{ACC(\(15^\circ\))} $\uparrow$ &  \textbf{ACC(\(10^\circ\))} $\uparrow$ & \textbf{MAE {[}\(^\circ\){]} $\downarrow$} \\ \midrule 
Block \(\times\) 3 (Expand = [2,2,2]) &  0.9  &  96.5        & 92.6                      & 2.89                \\ \midrule
Block \(\times\) 4 (Expand = [2,2,4,4])&  1.3  &  98.4        & 94.9                      & 2.87                \\ \midrule
\rowcolor[HTML]{C4D5EB} Block \(\times\) 5 (Expand = [2,2,4,4,8])& 1.8     & \textbf{98.9}     & \textbf{96.9}                      & \textbf{2.52}                \\ 
\multicolumn{1}{c|}{w/o BiMamba}& 1.4  & 97.1 & 93.6                      & 2.84                \\ 
\multicolumn{1}{c|}{w/o Skip}& 1.8     &  97.9     & 95.4                      & 2.65                \\ \bottomrule
\end{tabular}%
}
\end{table}

\begin{table}[t]
\centering
\caption{Performance comparison in ACC and MAE on simulated data. We test the models under the clean set and with -10 dB noise. The highlight is our proposed TF-Mamba model.}
\vspace{-2mm}
\label{tab:sim-table}
\resizebox{\columnwidth}{!}{%
\begin{tabular}{c|c|c|c|c|ccc}
\toprule
\multirow{1}{*}{\textbf{Methods}} &
  \multirow{1}{*}{\textbf{Params}}[M] $\downarrow$ &
  \multirow{1}{*}{\textbf{Clean}} &
  % \multicolumn{3}{c}{\textbf{Clean}} &
  % \multicolumn{2}{c}{\textbf{0 dB}} &
  % \multicolumn{3}{c}{\textbf{-10 dB}} \\ \cline{3-8} 
  \multicolumn{1}{c|}{\textbf{ACC(\(15^\circ\))} $\uparrow$} &
  \textbf{ACC(\(10^\circ\))} $\uparrow$ &
  \multicolumn{1}{c}{\textbf{MAE{[}\(^\circ\){]}} $\downarrow$ } &
  % \textbf{ACC(\(15^\circ\))} &
  % \multicolumn{1}{c}{\textbf{ACC(\(10^\circ\))}} &
  % \textbf{MAE{[}\(^\circ\){]}} 
  \\ \midrule

SELDnet & 0.8 & \multirow{6}{*}{\textbf{\cmark}} 
   & \multicolumn{1}{c|}{96.6} & 
  \multicolumn{1}{c|}{93.8} & 3.5
 
   \\ 

FN-SSL & 2.1 &
   & \multicolumn{1}{c|}{98.3} & 
  \multicolumn{1}{c|}{95.9} & 2.8
   \\ 
   
Cross3D & 5.6 &
   &  \multicolumn{1}{c|}{84.8} & 
  \multicolumn{1}{c|}{81.3} & 6.5

   \\ 
   
SE-ResNet & 10.2 &
   &  \multicolumn{1}{c|}{96.7} & 
  \multicolumn{1}{c|}{95.1} & 3.2

   \\ 
   
SALSA-Lite & 14.0 &
   & \multicolumn{1}{c|}{96.8} & 
  \multicolumn{1}{c|}{95.6} & 2.9

   \\ 
\rowcolor[HTML]{C4D5EB} \multicolumn{1}{c|}{TF-Mamba} &
  \multicolumn{1}{c|}{1.8}  & & \textbf{98.9} & 
  \textbf{96.9} &
   \textbf{2.5}
  \\ \midrule

  SELDnet & 0.8 & \multirow{6}{*}{\textbf{\xmark}}
   &   52.8
   &
  \multicolumn{1}{c|}{50.2} & 16.7

   \\ 

FN-SSL & 2.1 &
   &   71.6
   &
  \multicolumn{1}{c|}{68.4} & 8.9

   \\ 
   
Cross3D & 5.6 &
   &    38.8
   &
  \multicolumn{1}{c|}{35.4} & 26.3

   \\ 
   
SE-ResNet & 10.2 &
   &   67.7
   &
  \multicolumn{1}{c|}{64.6} & 10.8

   \\ 
   
SALSA-Lite & 14.0 &
   & 69.1
   & 
  \multicolumn{1}{c|}{65.7} & 10.3
 
   \\ 
\rowcolor[HTML]{C4D5EB} \multicolumn{1}{c|}{TF-Mamba} & 
  \multicolumn{1}{c|}{1.8} & & \textbf{74.6}
&
  \multicolumn{1}{c|}{\textbf{72.5}} &
  \multicolumn{1}{c}{\textbf{8.2}} 
  \\ \bottomrule
\end{tabular}%
}
\vspace{-4mm}
\end{table}

% Please add the following required packages to your document preamble:
% \usepackage{graphicx}
\begin{table}[t]
\centering
\caption{Performance comparison in ACC and MAE on LOCATA dataset. The highlight is our proposed TF-Mamba model.}
\vspace{-2mm}
\label{tab:locata-table}
\resizebox{0.9\columnwidth}{!}{%
\begin{tabular}{c|c|c|c|c}
\toprule
\textbf{Methods}                        & \textbf{Params[M]} $\downarrow$            & \textbf{ACC(\(15^\circ\))} $\uparrow$      & \textbf{ACC(\(10^\circ\))} $\uparrow$ & \textbf{MAE {[}\(^\circ\){]}} $\downarrow$                   \\ \midrule 
SELDnet    & 0.8 & 94.1 & 91.2 &  5.3\\ 
FN-SSL     & 2.1 & 96.7 & 93.3 & 3.6 \\ 
Cross3D    & 5.6 & 94.8 & 91.4 & 5.1 \\ 
SE-ResNet    & 10.2 & 94.0 & 90.7 & 6.5 \\ 
SALSA-Lite & 14.0 & 95.0 & 92.5 & 4.6 \\ 
\rowcolor[HTML]{C4D5EB} \multicolumn{1}{c|}{TF-Mamba} & \multicolumn{1}{c|}{1.8} & \multicolumn{1}{c|}{\textbf{97.2}} &         \textbf{94.3}        & \multicolumn{1}{c}{\textbf{3.2}} \\ \bottomrule
\end{tabular}%
}
\vspace{-4mm}
\end{table}

\section{Results and Analysis}
\label{sec:res}
\subsection{Ablation: Configuration for TF-Mamba}
The ablation study results for TF-Mamba, shown in Table~\ref{tab:abl-table}, highlight the importance of different components in the network. ``Block \(\times\) $N$" refers to stacking $N$ number of TF-Mamba blocks. As the number of blocks increases from 3 to 5, accuracy  (\(10^\circ\) error
tolerances) improves from 92.6\% to 96.9\%, showing that deeper networks help the model understand more complex localization information. The ``w/o BiMamba" variant, which replaces the bidirectional mamba with unidirectional, reduces accuracy to 93.6\%  (\(10^\circ\)) and 97.1\%  (\(15^\circ\)). This shows that processing both past and future information is important for achieving good performance. The ``w/o Skip" variant, which excludes skip connections, also leads to a small drop in accuracy and a higher error. This confirms that skip connections are helpful in keeping information flowing through the network. We use Block \(\times\) 5 version with BiMamba and skip connections in the following comparison experiments.
\subsection{Results on simulated data}
We report the results in two acoustic conditions of testing in Table~\ref{tab:sim-table} which are ``Clean" and ``SNR=-10 dB". The results demonstrate that TF-Mamba outperforms other methods in clean and noisy environments, highlighting its robustness in handling noise and reverberation. With the highest accuracy in  \(10^\circ\) (96.9\%),  \(15^\circ\) (98.9\%) and lowest MAE (2.5) in clean conditions, TF-Mamba outperforms FN-SSL by more effectively capturing long-range time and frequency dependencies, resulting in higher accuracy and greater robustness in challenging environments. This improvement demonstrates our contribution by validating the effectiveness of dual-dimension Mamba modeling. In contrast, methods like SALSA-Lite, which processes noisy IPD with magnitude spectrum, and Cross3D, which uses the SRP-PHAT spatial spectrum, show significant performance drops under noise, underscoring the advantage of TF-Mamba. 

\subsection{Results on real-world data}
The results on the LOCATA dataset are presented in Table~\ref{tab:locata-table}, where the models trained on simulated data are directly tested. We observe that TF-Mamba achieves the best performance, with the highest accuracy at both 15° (97.2\%) and 10° (94.3\%) error tolerances, and the lowest MAE (3.2). This dataset's relatively good acoustic conditions, which are almost noise-free and have an RT60 of 0.55s, allow all models to perform well. However, TF-Mamba's superior accuracy and lower error indicate its ability to better capture fine localization details compared to the other methods. Notably, methods like SE-ResNet and SALSA-Lite, which are more complex in terms of parameters, show higher MAE, while simpler models like SELDnet struggle with accuracy and MAE. TF-Mamba's approach also avoids the over-smoothing problem seen in other models during sudden sound source turns, maintaining strong localization performance in real-world conditions.
% \section{Acknowledgment}
% We wanted to extend our sincere thanks to Yabo Wang et.al. (The authors of the~\cite{fu1}) for their suggestions during my work on reproducing the FN-SSL model and other baseline models.
\section{Conclusion}
\label{sec:con}
In this paper, we introduced TF-Mamba, a novel architecture for SSL that leverages the Mamba to fuse time and frequency features effectively. By replacing traditional LSTM blocks with bidirectional Mamba, TF-Mamba enhances the modeling of audio sequence contexts while maintaining computational efficiency. Our experimental results, conducted on both simulated and real-world datasets, demonstrate that TF-Mamba significantly outperforms existing state-of-the-art methods. These findings highlight the proposed TF-Mamba in advancing SSL tasks, marking the first application of such models in this field.
% \section{Acknowledgement}
% The authors would like to thank Yabo Wang et. al. (The authors of FN-SSL~\cite{fu1}) for their valuable suggestions during this work on reproducing the FN-SSL model and other baseline models.
% \vfill\pagebreak
% \clearpage
% \ifinterspeechfinal
%      The Interspeech 2025 organisers
% \else
%      The authors
% \fi

\footnotesize
\bibliographystyle{IEEEtran}
\bibliography{mybib}

% Generated by IEEEtran.bst, version: 1.13 (2008/09/30)
\begin{thebibliography}{10}
\providecommand{\url}[1]{#1}
\csname url@samestyle\endcsname
\providecommand{\newblock}{\relax}
\providecommand{\bibinfo}[2]{#2}
\providecommand{\BIBentrySTDinterwordspacing}{\spaceskip=0pt\relax}
\providecommand{\BIBentryALTinterwordstretchfactor}{4}
\providecommand{\BIBentryALTinterwordspacing}{\spaceskip=\fontdimen2\font plus
\BIBentryALTinterwordstretchfactor\fontdimen3\font minus \fontdimen4\font\relax}
\providecommand{\BIBforeignlanguage}[2]{{%
\expandafter\ifx\csname l@#1\endcsname\relax
\typeout{** WARNING: IEEEtran.bst: No hyphenation pattern has been}%
\typeout{** loaded for the language `#1'. Using the pattern for}%
\typeout{** the default language instead.}%
\else
\language=\csname l@#1\endcsname
\fi
#2}}
\providecommand{\BIBdecl}{\relax}
\BIBdecl

\bibitem{ssl1}
P.-A. Grumiaux, S.~Kitić, L.~Girin, and A.~Guérin, ``{A survey of sound source localization with deep learning methods},'' \emph{The Journal of the Acoustical Society of America}, vol. 152, no.~1, pp. 107--151, 07 2022.

\bibitem{ss}
P.~Chiariotti, M.~Martarelli, and P.~Castellini, ``{Acoustic beamforming for noise source localization--Reviews, methodology and applications},'' \emph{Mechanical Systems and Signal Processing}, vol. 120, pp. 422--448, 2019.

\bibitem{tse}
Z.~Zhang, Y.~Xu, M.~Yu, S.-X. Zhang, L.~Chen, and D.~Yu, ``{ADL-MVDR: All deep learning MVDR beamformer for target speech separation},'' in \emph{Proc. IEEE International Conference on Acoustics, Speech and Signal Processing (ICASSP)}, 2021, pp. 6089--6093.

\bibitem{xiaoicassp}
Y.~Xiao and R.~K. Das, ``{Dual Knowledge Distillation for Efficient Sound Event Detection},'' in \emph{Proc. International Conference on Acoustics, Speech and Signal Processing Workshop (ICASSPW)}.\hskip 1em plus 0.5em minus 0.4em\relax IEEE, 2024, pp. 690--694.

\bibitem{asr}
H.-Y. Lee, J.-W. Cho, M.~Kim, and H.-M. Park, ``{DNN-based feature enhancement using DOA-constrained ICA for robust speech recognition},'' \emph{IEEE Signal Processing Letters}, vol.~23, no.~8, pp. 1091--1095, 2016.

\bibitem{se}
A.~Xenaki, J.~B{\"u}nsow~Boldt, and M.~Gr{\ae}sb{\o}ll~Christensen, ``Sound source localization and speech enhancement with sparse bayesian learning beamforming,'' \emph{The Journal of the Acoustical Society of America}, vol. 143, no.~6, pp. 3912--3921, 2018.

\bibitem{itd}
M.~Raspaud, H.~Viste, and G.~Evangelista, ``{Binaural source localization by joint estimation of ILD and ITD},'' \emph{IEEE Transactions on Audio, Speech, and Language Processing}, vol.~18, no.~1, pp. 68--77, 2009.

\bibitem{rtf}
W.~Zhang and B.~D. Rao, ``A two microphone-based approach for source localization of multiple speech sources,'' \emph{IEEE Transactions on Audio, Speech, and Language Processing}, vol.~18, no.~8, pp. 1913--1928, 2010.

\bibitem{srp}
R.~Schmidt, ``Multiple emitter location and signal parameter estimation,'' \emph{IEEE Transactions on Antennas and Propagation}, vol.~34, no.~3, pp. 276--280, 1986.

\bibitem{fu2}
T.~N.~T. Nguyen, D.~L. Jones, K.~N. Watcharasupat, H.~Phan, and W.-S. Gan, ``{{SALSA-Lite}: A fast and effective feature for polyphonic sound event localization and detection with microphone arrays},'' in \emph{Proc. IEEE International Conference on Acoustics, Speech and Signal Processing (ICASSP)}, 2022, pp. 716--720.

\bibitem{fu3}
A.~Politis, K.~Shimada, P.~Sudarsanam, S.~Adavanne, D.~Krause, Y.~Koyama, N.~Takahashi, S.~Takahashi, Y.~Mitsufuji, and T.~Virtanen, ``{STARSS22}: {A} dataset of spatial recordings of real scenes with spatiotemporal annotations of sound events,'' in \emph{Proc. the Detection and Classification of Acoustic Scenes and Events Workshop (DCASE)}, 2022, pp. 125--129.

\bibitem{cdoa}
Y.~Xiao and R.~K. Das, ``Where's that voice coming? {Continual} learning for sound source localization,'' in \emph{Proc. IEEE International Conference on Multimedia and Expo (ICME)}, 2024.

\bibitem{cnn1}
D.~Suvorov, R.~Zhukov, and G.~Dong, ``Deep residual network for sound source localization in the time domain,'' \emph{Journal of Engineering and Applied Sciences}, vol.~13, no.~13, pp. 5096--5104, 2018.

\bibitem{cnn2}
B.~Yang, H.~Liu, and X.~Li, ``{SRP-DNN}: Learning direct-path phase difference for multiple moving sound source localization,'' in \emph{Proc. IEEE International Conference on Acoustics, Speech and Signal Processing (ICASSP)}, 2022, pp. 721--725.

\bibitem{cnn3}
N.~Ma, T.~May, and G.~J. Brown, ``Exploiting deep neural networks and head movements for robust binaural localization of multiple sources in reverberant environments,'' \emph{IEEE Transactions on Audio, Speech, and Language Processing}, vol.~25, no.~12, pp. 2444--2453, 2017.

\bibitem{cnn4}
S.~Adavanne, A.~Politis, and T.~Virtanen, ``Direction of arrival estimation for multiple sound sources using convolutional recurrent neural network,'' in \emph{Proc. IEEE 26th European Signal Processing Conference (EUSIPCO)}, 2018, pp. 1462--1466.

\bibitem{fu1}
Y.~Wang, B.~Yang, and X.~Li, ``{FN-SSL}: Full-band and narrow-band fusion for sound source localization,'' in \emph{Proc. Interspeech}, 2023, pp. 3779--3783.

\bibitem{ssm1}
A.~Gu, K.~Goel, and C.~Re, ``Efficiently modeling long sequences with structured state spaces,'' in \emph{Proc. International Conference on Learning Representations (ICLR)}, 2021.

\bibitem{ssm2}
A.~Gu, K.~Goel, A.~Gupta, and C.~Re, ``On the parameterization and initialization of diagonal state space models,'' in \emph{Proc. Advances in Neural Information Processing Systems (NIPS)}, vol.~35, 2022, pp. 35\,971--35\,983.

\bibitem{visionmamba}
L.~Zhu, B.~Liao, Q.~Zhang, X.~Wang, W.~Liu, and X.~Wang, ``{Vision Mamba}: Efficient visual representation learning with bidirectional state space model,'' in \emph{Proc. International Conference on Machine Learning (ICML)}, 2024.

\bibitem{jamba}
X.~Zhang, J.~Ma, M.~Shahin, B.~Ahmed, and J.~Epps, ``Rethinking mamba in speech processing by self-supervised models,'' in \emph{IEEE International Conference on Acoustics, Speech and Signal Processing (ICASSP)}, 2025, pp. 1--5.

\bibitem{mambainspeech}
X.~Zhang, Q.~Zhang, H.~Liu, T.~Xiao, X.~Qian, B.~Ahmed, E.~Ambikairajah, H.~Li, and J.~Epps, ``{Mamba in Speech}: Towards an alternative to self-attention,'' \emph{IEEE Transactions on Audio, Speech and Language Processing}, vol.~33, pp. 1933--1948, 2025.

\bibitem{spmamba}
X.~Jiang, C.~Han, and N.~Mesgarani, ``Dual-path mamba: Short and long-term bidirectional selective structured state space models for speech separation,'' in \emph{IEEE International Conference on Acoustics, Speech and Signal Processing (ICASSP)}.\hskip 1em plus 0.5em minus 0.4em\relax IEEE, 2025, pp. 1--5.

\bibitem{xlsrmamba}
Y.~Xiao and R.~K. Das, ``{XLSR-Mamba:} a dual-column bidirectional state space model for spoofing attack detection,'' \emph{IEEE Signal Processing Letters}, vol.~32, pp. 1276--1280, 2025.

\bibitem{semamba}
R.~Chao, W.-H. Cheng, M.~La~Quatra, S.~M. Siniscalchi, C.-H.~H. Yang, S.-W. Fu, and Y.~Tsao, ``An investigation of incorporating mamba for speech enhancement,'' in \emph{IEEE Spoken Language Technology Workshop (SLT)}, 2024, pp. 302--308.

\bibitem{mamba}
A.~Gu and T.~Dao, ``Mamba: Linear-time sequence modeling with selective state spaces,'' in \emph{Proc. Conference on Language Modeling (COLM)}, 2024.

\bibitem{se1}
C.~Quan and X.~Li, ``Multichannel long-term streaming neural speech enhancement for static and moving speakers,'' \emph{IEEE Signal Processing Letters}, pp. 1--5, 2024.

\bibitem{audiomamba}
M.~H. Erol, A.~Senocak, J.~Feng, and J.~S. Chung, ``{Audio Mamba: Bidirectional State Space Model for Audio Representation Learning},'' \emph{IEEE Signal Processing Letters}, vol.~31, pp. 1933--1948, 2024.

\bibitem{rawbmamba}
Y.~Chen, J.~Yi, J.~Xue, C.~Wang, X.~Zhang, S.~Dong, S.~Zeng, J.~Tao, L.~Zhao, and C.~Fan, ``{RawBMamba}: End-to-end bidirectional state space model for audio deepfake detection,'' in \emph{Proc. Interspeech}, 2024, pp. 2720--2724.

\bibitem{mv}
D.~Diaz-Guerra, A.~Miguel, and J.~R. Beltran, ``{Robust sound source tracking using SRP-PHAT and 3D convolutional neural networks},'' \emph{IEEE Transactions on Audio, Speech, and Language Processing}, vol.~29, pp. 300--311, 2020.

\bibitem{seresnet}
J.~S. Kim, H.~J. Park, W.~Shin, and S.~W. Han, ``A robust framework for sound event localization and detection on real recordings,'' \emph{Tech. Rep.}, 2022.

\bibitem{seldnet}
S.~Adavanne, A.~Politis, J.~Nikunen, and T.~Virtanen, ``Sound event localization and detection of overlapping sources using convolutional recurrent neural networks,'' \emph{IEEE Journal of Selected Topics in Signal Processing}, vol.~13, no.~1, pp. 34--48, 2018.

\bibitem{salsa}
T.~N.~T. Nguyen, D.~L. Jones, K.~N. Watcharasupat, H.~Phan, and W.-S. Gan, ``{SALSA-Lite: A fast and effective feature for polyphonic sound event localization and detection with microphone arrays},'' in \emph{Proc. IEEE International Conference on Acoustics, Speech and Signal Processing (ICASSP)}, 2022, pp. 716--720.

\bibitem{densenet}
G.~Huang, Z.~Liu, L.~Van Der~Maaten, and K.~Q. Weinberger, ``Densely connected convolutional networks,'' in \emph{Proc. IEEE Conference on Computer Vision and Pattern Recognition (CVPR)}, 2017, pp. 4700--4708.

\bibitem{ssnet1}
W.~He, P.~Motlicek, and J.-M. Odobez, ``Neural network adaptation and data augmentation for multi-speaker direction-of-arrival estimation,'' \emph{IEEE Transactions on Audio, Speech, and Language Processing}, vol.~29, pp. 1303--1317, 2021.

\bibitem{librispeech}
V.~Panayotov, G.~Chen, D.~Povey, and S.~Khudanpur, ``Librispeech: an asr corpus based on public domain audio books,'' in \emph{Proc. IEEE International Conference on Acoustics, Speech and Signal Processing (ICASSP)}, 2015, pp. 5206--5210.

\bibitem{gpurir}
D.~Diaz-Guerra, A.~Miguel, and J.~R. Beltran, ``{gpuRIR: A python library for room impulse response simulation with GPU acceleration},'' \emph{Multimedia Tools and Applications}, vol.~80, no.~4, pp. 5653--5671, 2021.

\bibitem{NOISEX92}
A.~Varga and H.~J. Steeneken, ``{Assessment for automatic speech recognition: II. NOISEX-92: A database and an experiment to study the effect of additive noise on speech recognition systems},'' \emph{Speech communication}, vol.~12, no.~3, pp. 247--251, 1993.

\bibitem{generating}
E.~A. Habets, I.~Cohen, and S.~Gannot, ``Generating nonstationary multisensor signals under a spatial coherence constraint,'' \emph{The Journal of the Acoustical Society of America}, vol. 124, no.~5, pp. 2911--2917, 2008.

\bibitem{locata}
H.~W. L{\"o}llmann, C.~Evers, A.~Schmidt, H.~Mellmann, H.~Barfuss, P.~A. Naylor, and W.~Kellermann, ``{The LOCATA challenge data corpus for acoustic source localization and tracking},'' in \emph{Proc. IEEE Sensor Array and Multichannel Signal Processing Workshop (SAM)}, 2018, pp. 410--414.

\bibitem{woz}
A.~Brutti, L.~Cristoforetti, W.~Kellermann, L.~Marquardt, and M.~Omologo, ``{WOZ acoustic data collection for interactive TV},'' \emph{Language Resources and Evaluation}, vol.~44, pp. 205--219, 2010.

\end{thebibliography}

\end{document}